\documentclass[pre,nofootinbib]{revtex4}
\usepackage{amsfonts}
\usepackage{amsmath}
\usepackage{amssymb}
\usepackage{graphicx}

\input{tcilatex}

\begin{document}

\title{Statistical Mechanics of DNA-Mediated Colloidal Aggregation}
\author{Nicholas A. Licata, Alexei V. Tkachenko}
\affiliation{Department of Physics and Michigan Center for Theoretical Physics,
University of Michigan, \ 450 Church Str., Ann Arbor, Michigan 48109}

\begin{abstract}
We present a statistical mechanical model of aggregation in colloidal
systems with DNA mediated interactions. \ We obtain a general result for the
two-particle binding energy \ in terms of the hybridization free energy $%
\Delta G$ of DNA and two model dependent properties: the average number of
available DNA\ bridges $\left\langle N\right\rangle $ and the effective DNA\
conccentration $c_{eff}$. \ We calculate these parameters for a particular
DNA bridging scheme. \ The fraction of all the $n$-mers, including the
infinite aggregate, are shown to be universal functions of a single
parameter directly related to the two-particle binding energy. \ We
explicitly take into account the partial ergodicity of the problem resulting
from the slow DNA binding-unbinding dynamics, and introduce the concept of
angular localization of DNA\ linkers. \ In this way, we obtain a direct link
between DNA\ thermodynamics and the global aggregation and melting
properties in DNA-colloidal systems. \ The results of the theory are shown
to be in quantitative agreement with two recent experiments with particles
of micron and nanometer size. \ PACS numbers: 81.16.Dn, 82.20.Db, 68.65.-k,
87.14.Gg
\end{abstract}

\maketitle

\address{Department of Physics and Michigan Center for Theoretical Physics, \\
University of Michigan, 450 Church Street, Ann Arbor, Michigan 48109}

\section{Introduction\protect\bigskip}

In the past ten years, there have been a number of advances in experimental
assembly of nanoparticles with DNA-mediated interactions(\cite{rational},%
\cite{programmed},\cite{synthesis},\cite{supra},\cite{alivisatos},\cite%
{braun}). \ While this approach has a potential of generating highly
organized and sophisticated structures(\cite{morphology},\cite{licata}),
most of the studies report random aggregation of colloidal particles(\cite%
{chaikin},\cite{crocker}). \ Despite these shortcomings, the aggregation and
melting properties may provide important information for future development
of DNA-based self--assembly techniques. \ These results also have more
immediate implications. For instance, the observed sharp melting transition
is of particular interest for biosensor applications \cite{biotech}. \ For
these reasons the development of a statistical mechanical description of
these types of systems is of great importance. \ It should be noted that the
previous models of aggregation in colloidal-DNA systems were either
phenomenological or oversimplified lattice models(\cite{meltingprop},\cite%
{phasebehav},\cite{theory}), which gave only limited insight into the
physics of the phenomena. \ 

In this paper, we develop a theory of reversible aggregation and melting in
colloidal-DNA systems, starting from the known thermodynamic parameters of
DNA (i.e. hybridization free energy $\Delta G$), and geometric properties of
DNA-particle complexes. \ The output of our theory is the relative abundance
of the various colloidal structures formed (dimers, trimers, etc.) as a
function of temperature, as well as the temperature at which a transition to
an infinite aggregate occurs. \ The theory provides a direct link between
DNA\ microscopics and experimentally observed morphological and thermal
properties of the system. \ It should be noted that the hybridization free
energy $\Delta G$ depends not only on the DNA\ nucleotide sequence, but also
on the salt concentration and the concentration of DNA\ linker strands
tethered on the particle surface\cite{tetherhyb}. \ In this paper $\Delta G$
values refer to hybridization between DNA free in solution. \ 

\begin{figure}[h]
{\includegraphics[width=3.9332in,height=2.9603in]{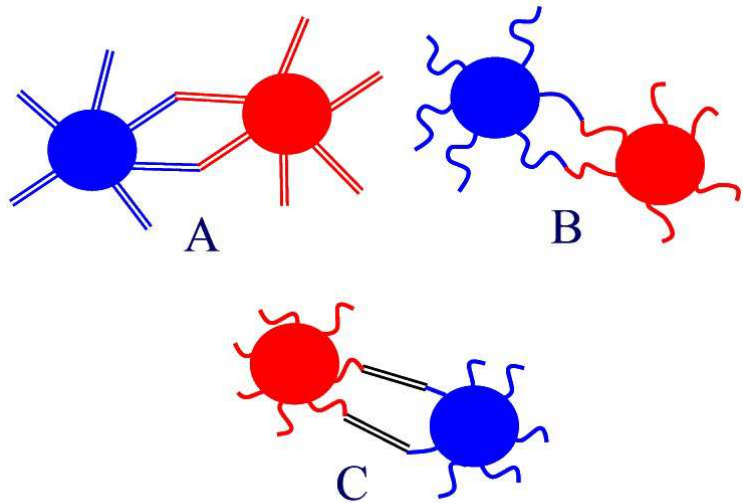}}
\caption{(Color Online) Graphical depiction of various schemes for DNA\
bridging. \ A) A freely-jointed, rigid bridge constructed from complementary
linker DNA. \ B) A flexible bridge can be constructed using complementary
linker DNA. \ C) A rigid bridge constructed from short, flexible linker DNA
and a long, rigid linker. \ }
\label{schemes}
\end{figure}

In a generic experimental setup, \ particles are grafted with DNA\ linker
sequences which determine the particle type($A$ or $B$). \ In this paper we
will restrict our attention to a binary system\footnote{%
This restriction to binary systems is consistent with the current
experimental approach. \ In a recent work we demonstrated that if each
particle has a \textit{unique} linker sequence, one might be able to
programmably self-assemble nanoparticle clusters of desired geometry\cite%
{licata}.}. \ These linkers may be flexible or rigid. \ A selective,
attractive potential between particles of type $A$ and $B$ can then be
turned on by joining the linkers to form a DNA\ bridge. \ This DNA bridge
can be constructed directly if the particle linker sequences are chosen to
have complementary ends. \ Alternatively, the DNA bridge can be constructed
with the help of an additional linker DNA. \ This additional linker is
designed to have one end sequence complementary to the linker sequence of
type $A$ particles, and the other end complementary to type $B$. \ The
properties of the DNA bridge formed will depend on the hybridization
scheme(see figure \ref{schemes}). \ 

The plan for the paper is as follows. \ In section II.A we provide a
description of the problem. \ In section II.B we determine the bridging
probability for the formation of a DNA\ bridge between two colloids,
assuming the known thermodynamic parameters of DNA(hybridization free energy 
$\Delta G$). \ Using this bridging probability as input, in section II.C we
calculate the effective binding free energy $\epsilon_{AB}$ for the
formation of a dimer. \ Sections III.A-III.C establish the connection
between the theory and the experimentally determined melting profile $f(T)$,
the fraction of unbound particles as a function of temperature. \ In
particular, we demonstrate how knowledge of $\epsilon_{AB}$ can be used to
determine this profile, including the effects of particle aggregation. \ In
Section III.D the theory is compared with two recent experiments detailing
the reversible aggregation of colloids with DNA-mediated attraction\cite%
{meltingprop},\cite{chaikin}. \ The main results of the model are summarized
in section IV. \ 

\section{DNA-mediated interactions}

\subsection{Description of the problem}

We consider particles of type $A$ and $B$ which form reversible $AB$ bonds
as a result of DNA hybridization. \ The task at hand is to determine the
relative abundance of the various colloidal structures that form as a
function of temperature. \ From this information we can determine which
factors affect the melting and aggregation properties in DNA-colloidal
assemblies. \ To do so we must determine the binding free energy for all of
the possible phases(monomer, dimer, ..., infinite aggregate), and then apply
the rules for thermodynamic equilibrium. \ As we will see in section III,
these binding free energies can all be simply related to $\epsilon_{AB}$,
the binding free energy for the formation of a dimer. \ Our task is thus
reduced to determining $\epsilon _{AB}$ from the thermodynamic parameters of
DNA and structural properties of the DNA\ linkers. \ In our statistical
mechanical framework, $\epsilon_{AB}$ is calculated from the model partition
function, taking into account the appropriate ensemble averaging for the
non-ergodic degrees of freedom. \ The result is related to the bridging
probability for a pair of linkers. \ By considering the specific properties
of the DNA bridge that forms, the bridging probability can be related to the
hybridization free energy $\Delta G$ of the DNA. \ In this way, we obtain a
direct link between DNA\ thermodynamics and the global aggregation and
melting properties in colloidal-DNA systems. \ 

\subsection{Bridging probability}

To begin we relate the hybridization free energy $\Delta G$ for the DNA in
solution to the bridging probability for a pair of linkers. \ This bridging
probability is defined as the ratio $\frac{P_{bound}}{P_{free}}$, with $%
P_{bound}$ the probability that the pair of linkers have hybridized to form
a DNA bridge, and $P_{free}$ the probability that they are unbound. \ This
ratio is directly related to the free energy difference of the bound and
unbound states of the linkers $\Delta \widetilde{G}$ (throughout this paper
we will use units with $k_{B}=1$): \ 
\begin{align}
\frac{P_{bound}}{P_{free}}& =\exp \left[ \frac{-\Delta \widetilde{G}}{T}%
\right] =\frac{c_{eff}}{c_{o}}\exp \left[ \frac{-\Delta G}{T}\right] \\
c_{eff}& =\frac{\int P(\mathbf{r}_{1},\mathbf{r})P(\mathbf{r}_{2},\mathbf{r}%
)d^{3}\mathbf{r}}{\left( \int P(\mathbf{r},\mathbf{r}^{\prime })d^{3}\mathbf{%
r}\right) ^{2}}  \label{ceff}
\end{align}%
Here $c_{o}=1M$ is a reference concentration. \ $P(\mathbf{r},\mathbf{r}%
^{\prime })$ is the probability distribution function for the linker chain
which starts at $\mathbf{r}^{\prime }$ and ends at $\mathbf{r}$. The
effective concentration $c_{eff}$ is a measure of the change in
conformational entropy of the linker DNA as a result of hybridization. \ It
will depend on the properties of the linker DNA(ex: flexible vs. rigid), and
the scheme for DNA\ bridging(ex: hybridization of complementary ends vs.
hybridization mediated by an additional linker). $\ c_{eff}$ is the
concentration of free DNA which would have the same hybridization
probability as the grafted linkers in our problem. \ As discussed in section
III.D, the DNA linker grafting density also plays an important role in
determining the possible linker configurations and hence $c_{eff}$. \ 

\begin{figure}[h]
{\includegraphics[width=4.2774in,height=3.2188in]{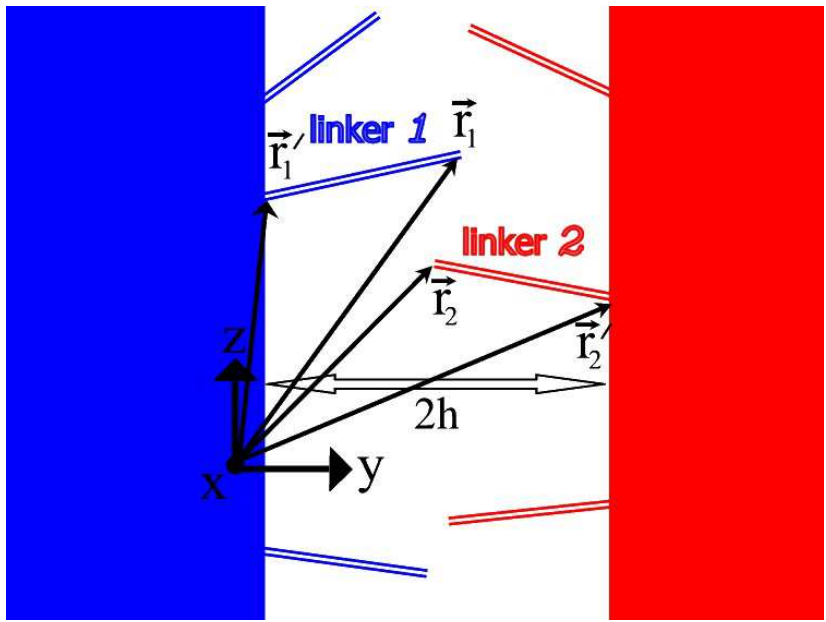}}
\caption{(Color Online) The statistical weight of a bound state is
calculated by determining the number of hybridized configurations for two
complementary linker chains relative to the number of unhybridized
configurations. \ }
\label{setup}
\end{figure}

Assuming that the size of the linkers is much smaller than the particle
radius $R$, we first consider the problem in a planar geometry. Let the two
linkers be attached to two parallel planar surfaces separated by a distance $%
2h$. \ Referring to figure \ref{setup} we see that $\mathbf{r}^{\prime}$ is
the location where the linker DNA is grafted onto the particle surface, and $%
\mathbf{r}$ is the position of the free end. \ \ 

In this paper we consider hybridization by complementary, rigid linker DNA
(scheme A in Figure \ref{schemes}). \ This scheme is particularly
interesting since it is directly related to several recent experiments \cite%
{chaikin},\cite{meltingprop}. \ In a future work we will address other
hybridization schemes. \ We assume that $L<L_{p}$ and $L\ll R$, where $%
L_{p}\simeq50nm$ is the persistence length of ds DNA and $L$ is the ds
linker DNA length. In this regime, the linker chains can be treated as rigid
rods tethered on a planar surface. \ The interaction is assumed to be
point-like, in which a small fraction $\Delta/L$\ of the linker bases
hybridize.

\begin{figure}[h]
{\includegraphics[width=4.2774in,height=3.2188in]{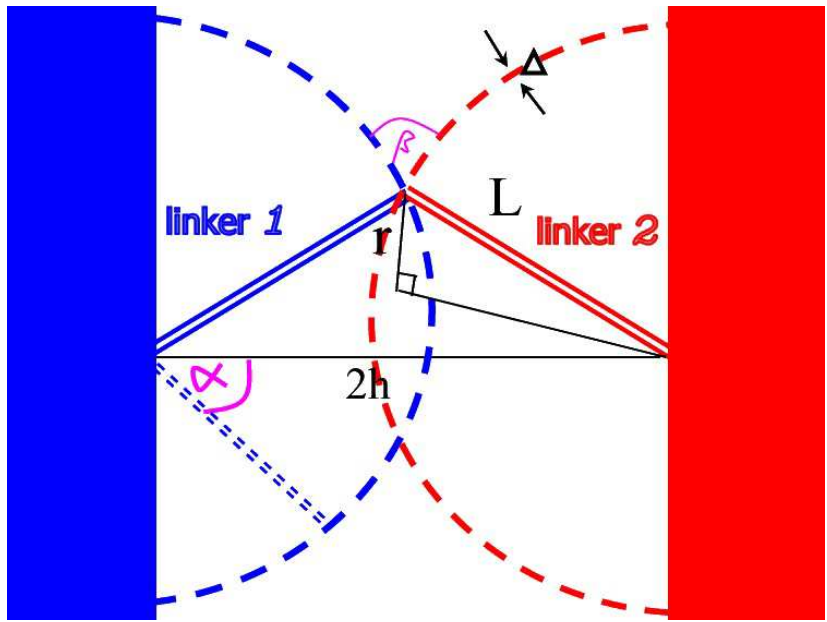}}
\caption{(Color Online) Cross-sectional view of the hybridization of two
complementary rigid linker DNA. \ The effective concentration is calculated
in a planar approximation to the particle surface. \ }
\label{hybridization}
\end{figure}

We can calculate the effective concentration by noting that the overlap
integral in Eq. (\ref{ceff}) is proportional to the volume of intersection
of two spherical shells (red and blue circles in Figure \ref{hybridization})
: \ 
\begin{equation}
c_{eff}=\frac{2\pi rA}{\left( 2\pi L^{2}\Delta\right) ^{2}}=\frac {%
\Theta\left( L-\left\vert \mathbf{r}_{1}^{\prime}-\mathbf{r}_{2}^{\prime
}\right\vert /2\right) }{2\pi^{2}L^{2}\left\vert \mathbf{r}_{1}^{\prime }-%
\mathbf{r}_{2}^{\prime}\right\vert },
\end{equation}
here $A=\Delta^{2}/\sin\beta\ $and $r=\sqrt{L^{2}-\left\vert \mathbf{r}
_{1}^{\prime}-\mathbf{r}_{2}^{\prime}\right\vert ^{2}/4}$ (see notations in
Figure \ref{hybridization}). We have used the fact that $\cos\beta
/2=\left\vert \mathbf{r}_{1}^{\prime}-\mathbf{r}_{2}^{\prime}\right\vert /2L$%
. $\ c_{eff}$ and the binding probability are largest when the linkers are
grafted right in front of each other, i.e. when $\left\vert \mathbf{r}
_{1}^{\prime}-\mathbf{r}_{2}^{\prime}\right\vert \sim2h$. \ By taking the
limit $h\approx L$\ we arrive at the following result for the corresponding
"bridging" free energy \ 
\begin{equation}
\Delta\widetilde{G}_{A}\approx\Delta G_{A}+T\log\left[ 4\pi L^{3} c_{o}%
\right] .  \label{rigidbridge}
\end{equation}

This free energy remains nearly constant for any pair of linkers, as long as
they can be connected in principle, i.e. $\left\vert \mathbf{r}_{1}^{\prime
}-\mathbf{r}_{2}^{\prime}\right\vert <2L$. This limits the maximum lateral
displacement of the linkers: \textbf{\ }$r_{\bot}<2\sqrt{L^{2}-h^{2}}$, and
therefore sets the \textit{effective cross-section} of the interaction: 
\begin{equation}
a=\mathbf{\ }\pi r_{\bot}^{2}=4\pi\left( L^{2}-h^{2}\right)
\label{rigidarea1}
\end{equation}

\qquad\qquad

\subsection{Effective binding free energy}

\qquad

\ We now proceed with the calculation of the effective free energy $\epsilon
_{AB}$ , which is associated with the formation of a dimer from a pair of
free particles, $A$ and $B$. \ Since the DNA coverage on the particle
surface is not uniform, this free energy, and the corresponding partition
function $Z$, would in principle depend on the orientations of the particles
with respect to the line connecting their centers. The equilibrium binding
free energy would correspond to the canonical ensemble of all possible
orientations, i.e. $\epsilon _{AB}=-T\log 4\pi \left\langle Z\right\rangle $%
. However, this equilibrium can only be achieved after a very long time,
when the particle pair samples all possible binding configurations, or at
least their representative subset. The real situation is different. After
the first DNA-mediated bridge is created the particle pair can still explore
the configurational space by rotating about this contact point. However,
after the formation of two or more DNA bridges (at certain relative
orientation of the particles), the further exploration requires multiple
breaking and reconnecting of the DNA links, which is a very slow process. We
conclude that the system is ergodic with respect to the various
conformations of the linker DNA for fixed orientations of the particles, but
the orientations themselves are \textit{non-ergodic variables}. The only
exceptions are the single-bridge states: the system quickly relaxes to a
more favorable orientational state (unless the DNA coverage is extremely
low, and finding a second contact is very hard). \ \ If $N$ denotes the
number of DNA\ bridges constituting the $AB$ bond, the appropriate
expression for $\epsilon _{AB}$ in this partially ergodic regime is the
so-called component averaged free energy(\cite{glasses},\cite%
{brokenergodicity}): 
\begin{equation}
\epsilon _{AB}=-T\left\langle \log Z\right\rangle _{N\geq 2}
\end{equation}

Each DNA bridge between particles can be either open or closed. \ 
\begin{equation}
Z_{bridge}=1+\exp\left( -\frac{\Delta\widetilde{G}(h^{\prime},\mathbf{r}
_{1}^{\prime}-\mathbf{r}_{2}^{\prime})}{T}\right)
\end{equation}
Here $\mathbf{r}_{i}^{\prime}$ is the 2D position where the bridge is
grafted onto surface $i$. \ We now consider a generic case when the
interaction free energy $\Delta\widetilde{G}$ depends on the separation
between planar surfaces $2h^{\prime}$, and the separation of grafting points 
$\mathbf{r}_{1}^{\prime }-\mathbf{r}_{2}^{\prime}$, without assumption of a
particular bridging scheme. \ If the probability for bridge formation is
small, two DNA linkers on the same surface will not compete for
complementary linkers. \ In this regime the free energy can be calculated by
summing over the contribution from each bridge that forms between dimers. \ 
\begin{equation}
F=-T\sum_{i}\sum_{j}\log\left[ 1+\exp\left( -\frac{\Delta\widetilde {G}%
(h^{\prime},\mathbf{r}_{i}^{\prime}-\mathbf{r}_{j}^{\prime})}{T}\right) %
\right]
\end{equation}
We convert the summation to integration by introducing the linker areal
grafting density $\sigma$. \ 
\begin{equation}
F=-T\int\int\sigma_{1}(\mathbf{r}_{1}^{\prime})\sigma_{2}(\mathbf{r}
_{2}^{\prime})\log\left[ 1+\exp\left( -\frac{\Delta\widetilde{G}(h^{\prime },%
\mathbf{r}_{1}^{\prime}-\mathbf{r}_{2}^{\prime})}{T}\right) \right] d^{2}%
\mathbf{r}_{1}^{\prime}d^{2}\mathbf{r}_{2}^{\prime}
\end{equation}
Changing variables to $\mathbf{\Delta r}=\mathbf{r}_{1}^{\prime} -\mathbf{r}%
_{2}^{\prime}$ and $\mathbf{\rho}=\left( \mathbf{r}_{1}^{\prime }+\mathbf{r}%
_{2}^{\prime}\right) /2$, we can reintroduce the notion of a bridging
cross-section $a(h^{\prime})$, this time in a model-independent manner: \ 
\begin{equation}
a(h^{\prime})\log\left[ 1+\exp\left( -\frac{\Delta\widetilde{G}
_{o}(h^{\prime})}{T}\right) \right] \equiv \int d^{2} \mathbf{%
\Delta r}\log\left[ 1+\exp\left( -\frac{\Delta\widetilde {G}(h^{\prime},%
\mathbf{\Delta r})}{T}\right) \right]  \label{area}
\end{equation}
\ Here $\Delta\widetilde{G}_{o}(h^{\prime})\equiv\Delta\widetilde{G}
(h^{\prime},\mathbf{\Delta r}=0)$ is the minimum free energy with respect to
the separation between grafting points $\mathbf{\Delta r}$.  We can
now write the free energy: 
\begin{equation}
F=-T\int\sigma_{1}(\overrightarrow{\mathbf{\rho}})\sigma_{2}(%
\overrightarrow {\mathbf{\rho}})a(h^{\prime})\log\left[ 1+\exp\left( -\frac{%
\Delta \widetilde{G}_{o}(h^{\prime})}{T}\right) \right] d^{2}\mathbf{\rho}
\end{equation}
We now convert from the planar geometry to the spherical particle geometry
using the Derjaguin approximation\cite{intermolecular}. \ 
\begin{align}
d^{2}\mathbf{\rho} & =\rho d\rho d\phi \\
h^{\prime} & =h+\frac{\rho^{2}}{2R}
\end{align}
Let $\Delta\widetilde{G}_{\ast}$ \ be the minimal value of the bridging free
energy. \ \ Then the result for $F$ can be rewritten as:

\bigskip 
\begin{equation}
F=-TN\log\left[ 1+\exp\left( -\frac{\Delta\widetilde{G}_{\ast}}{T}\right) %
\right]
\end{equation}
Here $N$ has a physical meaning as the number of potential bridges for given
relative positions and orientations of the particles: 
\begin{equation}
N\equiv\int\sigma_{1}(\overrightarrow{\mathbf{\rho}})\sigma_{2} (%
\overrightarrow{\mathbf{\rho}})a(h^{\prime})\left( \frac{\log\left[
1+\exp\left( -\Delta\widetilde{G}_{o}(h^{\prime})/T\right) \right] } {\log%
\left[ 1+\exp\left( -\Delta\widetilde{G}_{\ast}/T\right) \right] }\right)
d^{2}\mathbf{\rho}
\end{equation}

One can calculate the average value of $N$ in terms of the average grafting
density, $\sigma =\left\langle \sigma _{1}\right\rangle =\left\langle \sigma
_{2}\right\rangle :$ \ 
\begin{equation}
\left\langle N\right\rangle \equiv 2\pi R\sigma ^{2}\int a(h^{\prime
})\left( \frac{\log \left[ 1+\exp \left( -\Delta \widetilde{G}_{o}(h^{\prime
})/T\right) \right] }{\log \left[ 1+\exp \left( -\Delta \widetilde{G}_{\ast
}/T\right) \right] }\right) dh^{\prime }  \label{nbar}
\end{equation}%
In a generic case of randomly grafted linkers, $\left\langle N\right\rangle $
completely defines the overall distribution function of $N$, which must have
a Poisson form: $P(N)=\frac{\left\langle N\right\rangle ^{N}e^{-\left\langle
N\right\rangle }}{N!}$. \ The average number of bridges $\left\langle
N\right\rangle $ between two particles depends on both the DNA linker
grafting density $\sigma $ and the bridging probability determined from $%
\Delta \widetilde{G}$. \ 

The free energy for the formation of a dimer $\epsilon_{AB}=\left\langle
F\right\rangle _{2+}-T\log\Omega$. \ The second term is the entropic
contribution to the free energy, which comes from integration over the
orientational and translational degrees of freedom of the second particle. \
Because the system is not ergodic in these degrees of freedom, the
accessible phase space $\Omega$ will be reduced by a factor of $P_{2+}$. $\
P_{2+}$ is the probability that there are at least two DNA bridges between
the particles. \ In terms of the average number of bridges $\left\langle
N\right\rangle $ between particles, we have the following relations: \ 
\begin{align}
P_{2+} & =1-(1+\left\langle N\right\rangle )e^{-\left\langle N\right\rangle }
\\
\left\langle N\right\rangle _{2+} & =\frac{\left\langle N\right\rangle
\left( 1-e^{-\left\langle N\right\rangle }\right) }{P_{2+}}
\end{align}
\begin{equation}
\epsilon_{AB}=-T\left\{ \left\langle N\right\rangle _{2+}\log\left[
1+\exp\left( -\frac{\Delta\widetilde{G}_{\ast}}{T}\right) \right] +\log\left[
P_{2+}4\pi\delta(2R)^{2}c_{o}\right] \right\}
\end{equation}
Here $\delta$ is the localization length of the $AB$ bond, which comes from
integrating the partition function over the radial distance between
particles. \qquad

\bigskip We now can calculate $\left\langle N\right\rangle $ for the case of
freely-jointed rigid bridging considered earlier (i.e. for scheme A). \ In a
previous section we provided a direct calculation of the interaction free
energy, $\Delta\widetilde{G}_{o}\left( h\right) \approx const=\Delta 
\widetilde{G}_{\ast}$ (eq.\ref{rigidbridge}), and bridging cross-section, $%
a(h^{\prime})=4\pi(L^{2}-h^{\prime2})$. Applying eq. \ref{nbar} we arrive
immediately at the following result. \ 
\begin{equation}
\left\langle N\right\rangle =8\pi^{2}\sigma^{2}R\int\limits_{0}^{L}
(L^{2}-h^{\prime2})dh^{\prime}=\frac{16\pi^{2}\sigma^{2}RL^{3}}{3}
\end{equation}
Having determined the free energy, we are now in a position to determine the
melting properties for DNA colloidal assemblies. \ 

\section{Aggregation and melting in colloidal-DNA systems}

At this stage of the paper we have calculated the binding free energy $%
\epsilon_{AB}$ for an $AB$ pair, starting with the thermodynamic parameters
of DNA (hybridization free energy $\Delta G$). \ In this section of the
paper we establish the connection between that result and the experimentally
observable morphological behavior of a large system. One of the ways to
characterize the system is to study its melting profile $f(T)$, which is the
fraction of unbound particles as a function of temperature. \ To determine
the profile we calculate the chemical potential for each phase(monomer,
dimer, etc.) and apply the thermodynamic rules for phase equilibrium. \ We
will demonstrate how the single binding free energy $\epsilon_{AB}$ can be
used to determine the contribution of each phase to the melting profile,
including the effects of aggregation. \ 

\subsection{Dimer formation}

To begin we discuss the formation of dimers via the reaction $%
A+B\rightleftharpoons AB$. \ We can express the chemical potential of the $%
i^{th}$ species $\mu_{i}$ in terms of the particle concentrations $c_{i}=%
\frac{N_{i}}{V}$. \ 
\begin{align}
\mu_{A} & =T\log\left( c_{A}\right) \\
\mu_{B} & =T\log\left( c_{B}\right) \\
\mu_{AB} & =T\log\left( c_{AB}\right) +\epsilon_{AB}
\end{align}
Here $\epsilon_{AB}$ is the binding free energy for the formation of a
dimer. \ In terms of the potential $V(r)$ between $A$ and $B$ type particles
we have: 
\begin{equation}
\epsilon_{AB}=-T\log\left[ 4\pi(2R)^{2}c_{o}\int dr\exp\left( -\frac {V(r)}{T%
}\right) \right]  \label{dnapotential}
\end{equation}

In this section we are not particularly concerned with the specific form of
the DNA-induced potential $V(r)$, having already determined $\epsilon_{AB} $%
\ in the previous section. \ We simply note that the prefactor $4\pi(2R)^{2} 
$ arises since the interaction is assumed to be isotropic, with $R$ the
particle radius. \ Equilibrating the chemical potential of the various
particle species, we obtain the condition for chemical equilibrium. $\ $ 
\begin{equation}
\mu_{A}+\mu_{B}=\mu_{AB}
\end{equation}
The result is a relationship between the concentration of dimers and
monomers. \ \ 
\begin{equation}
c_{AB}=\frac{c_{A}c_{B}}{c_{o}}\exp\left[ \frac{-\epsilon_{AB}}{T}\right]
\end{equation}
The overall concentration of particles in monomers and dimers must not
differ from the initial concentration. \ 
\begin{align}
c_{A}^{i} & =c_{A}+c_{AB} \\
c_{B}^{i} & =c_{B}+c_{AB}
\end{align}
If the system is prepared at equal concentration, $c_{A}^{i}=c_{B}^{i} =%
\frac{1}{2}c_{tot}$, subtracting the two equations we see that $c_{A}
=c_{B}\equiv c $. \ Written in terms of the fraction of unbound particles $f=%
\frac{c}{\frac{1}{2}c_{tot}}$ we have a quadratic equation for the unbound
fraction. \ 
\begin{equation}
1=f+\exp\left[ \frac{-\widetilde{\epsilon}_{AB}}{T}\right] f^{2}
\end{equation}
To simplify we have defined an effective free energy $\widetilde{\epsilon }%
_{AB}$ for the formation of a dimer. \ 
\begin{equation}
\widetilde{\epsilon}_{AB}=\epsilon_{AB}-T\log\left[ \frac{c_{tot}}{2c_{o} }%
\right]
\end{equation}
The solution for the fraction of unbound particles as a function of
temperature is simply: 
\begin{equation}
f=\frac{-1+\sqrt{1+4\exp\left[ \frac{-\widetilde{\epsilon}_{AB}}{T}\right] }%
}{2\exp\left[ \frac{-\widetilde{\epsilon}_{AB}}{T}\right] }
\end{equation}
\qquad\qquad

Previous studies\cite{chaikin} only included the dimer contribution to the
melting properties of DNA colloidal assemblies. \ With the basic formalism
at hand, we can now extend the preceding analysis to include the
contribution of trimers and tetramers. \ 

\subsection{Trimers and tetramers}

Now consider the formation of a trimer via $2A+B\rightleftharpoons ABA$. \
The chemical potential is slightly different in this case. \ 
\begin{equation}
\mu_{ABA}=T\log\left( c_{ABA}\right) +\epsilon_{ABA}
\end{equation}
Taking into account that there are now two $AB$ bonds in the structure, one
might conclude that $\epsilon_{ABA}=2\epsilon_{AB}$. \ This is not quite
correct, since there is a reduction in solid angle available to the third
particle. \ To form a trimer, an $AB$ bond forms first, which contributes $%
\epsilon_{AB}$ to $\epsilon_{ABA}$. \ Some simple geometry shows that the
remaining $A$ particle only has $3\pi$ steradians of possible bonding sites
to particle $B$. \ Making this change in the prefactor of eq. \ref%
{dnapotential}, one can see that the second bond contributes $%
\epsilon_{AB}-T\log\left( \frac{3}{4}\right) $ to $\epsilon_{ABA}$. \ 
\begin{equation}
\epsilon_{ABA}=2\epsilon_{AB}-T\log\left( \frac{3}{4}\right)
\end{equation}
The equation for chemical equilibrium can once again be expressed in terms
of the particle concentrations. \ 
\begin{align}
2\mu_{A}+\mu_{B} & =\mu_{ABA} \\
c_{ABA} & =\frac{3}{4}\frac{c_{A}^{2}c_{B}}{c_{o}^{2}}\exp\left[ \frac{%
-2\epsilon_{AB}}{T}\right]
\end{align}
To include the trimer contribution, we note that there are two possible
varieties, with $\epsilon_{ABA}=\epsilon_{BAB}$. \ 

\begin{align}
c_{A}^{i} & =c_{A}+c_{AB}+2c_{ABA}+c_{BAB} \\
c_{B}^{i} & =c_{B}+c_{AB}+c_{ABA}+2c_{BAB}
\end{align}
Following the same line of reasoning as before, the resulting equation for
the unbound fraction $f$ is: 
\begin{equation}
1=f+\exp\left[ \frac{-\widetilde{\epsilon}_{AB}}{T}\right] f^{2}+\frac{9} {4}%
\exp\left[ \frac{-2\widetilde{\epsilon}_{AB}}{T}\right] f^{3}
\end{equation}

For tetramers we will follow the same general reasoning, however in this
case there are two different structure types. \ The reaction $%
2A+2B\rightleftharpoons ABAB$ results in the formation of string like
structures. 
\begin{equation}
\mu_{ABAB}=T\log\left( c_{ABAB}\right) +\epsilon_{ABAB}
\end{equation}
As in the trimer case, the last particle has $3\pi$ steradians of possible
bonding sites, and contributes $\epsilon_{AB}-T\log\left( \frac{3}{4}\right) 
$ to $\epsilon_{ABAB}$. \ 
\begin{align}
\epsilon_{ABAB} & =3\epsilon_{AB}-T\log\left[ \left( \frac{3}{4}\right) ^{2}%
\right] \\
2\mu_{A}+2\mu_{B} & =\mu_{ABAB} \\
c_{ABAB} & =\left( \frac{3}{4}\right) ^{2}\frac{c_{A}^{2}c_{B}^{2}} {%
c_{o}^{3}}\exp\left[ \frac{-3\epsilon_{AB}}{T}\right]
\end{align}
If an $A$ type particle approaches a trimer of variety $ABA$, a branched
structure can result. \ The reaction $3A+B\rightleftharpoons AAAB$ results
in the formation of these branched structures. \ 
\begin{equation}
\mu_{AAAB}=T\log\left( c_{AAAB}\right) +\epsilon_{AAAB}
\end{equation}
For the branched case, the last particle has approximately $2\pi$ steradians
of possible bonding sites, and contributes $\epsilon_{AB}-T\log\left( \frac{1%
}{2}\right) $ to $\epsilon_{AAAB}$. \ 
\begin{align}
\epsilon_{AAAB} & =3\epsilon_{AB}-T\log\left( \frac{3}{8}\right) \\
3\mu_{A}+\mu_{B} & =\mu_{AAAB} \\
c_{AAAB} & =\left( \frac{3}{8}\right) \frac{c_{A}^{3}c_{B}}{c_{o}^{3}} \exp%
\left[ \frac{-3\epsilon_{AB}}{T}\right]
\end{align}
To include all of the tetramer contributions, note that there are two
branched varieties, with $\epsilon_{AAAB}=\epsilon_{BBBA}$. \ Finally we
impose the constraint that the initial particle concentrations do not differ
from the concentration of all the n-mers, for n=1,2,3,4. \ 
\begin{align}
c_{A}^{i} & =c_{A}+c_{AB}+2c_{ABA}+c_{BAB}+2c_{ABAB}+3c_{AAAB}+c_{BBBA} \\
c_{B}^{i} & =c_{B}+c_{AB}+c_{ABA}+2c_{BAB}+2c_{ABAB}+c_{AAAB}+3c_{BBBA}
\end{align}
The final result is an equation for the unbound fraction $f$ expressed
entirely in terms of the effective free energy $\widetilde{\epsilon}_{AB}$
of a dimer. \ 
\begin{equation}
1=f+\exp\left[ \frac{-\widetilde{\epsilon}_{AB}}{T}\right] f^{2}+\frac{9} {4}%
\exp\left[ \frac{-2\widetilde{\epsilon}_{AB}}{T}\right] f^{3}+\frac {21}{8}%
\exp\left[ \frac{-3\widetilde{\epsilon}_{AB}}{T}\right] f^{4}
\label{polynomial}
\end{equation}

For high temperatures, the melting profile is governed by the solution to
this polynomial equation for $f$. \ For temperatures below the melting point
we expect to find particles in large extended clusters. \ We now proceed to
calculate the equilibrium condition between monomers in solution and the
aggregate. \ 

\subsection{Reversible sol-gel transition}

To understand the basic structure of the aggregate, we simply note that
there are many DNA\ attached to each particle. \ This gives rise to
branching, as in the discussion of possible tetramer structures. \ Since the
DNA\ which mediate the interaction are grafted onto the particle surface,
once two particles are bound, the relative orientation of the pair is
essentially fixed. \ The resulting aggregate is a tree-like structure, and
the transition to an infinite aggregate at low temperatures is analogous to
the sol-gel transition in branched polymers\cite{scaling}. \ 
\begin{figure}[h]
\begin{center}
\includegraphics[
height=3.2206in,
width=4.28in
]{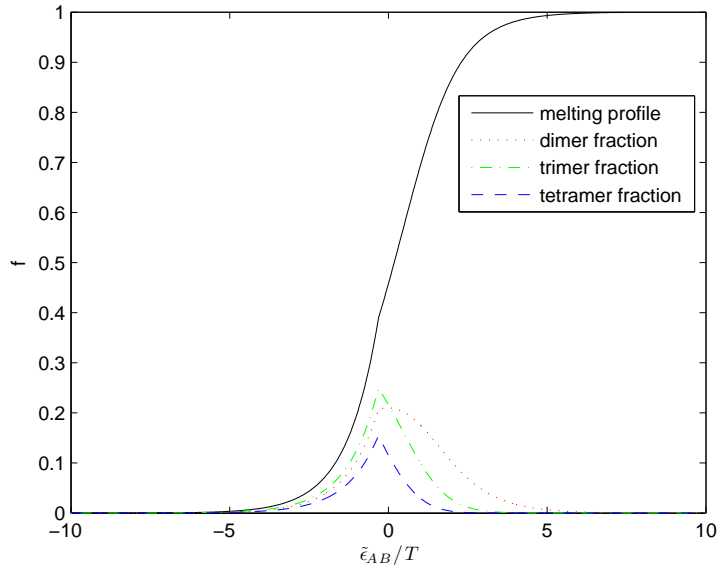}
\end{center}
\caption{(Color Online) The actual unbound fraction $f$\ is the
concatenation of the aggregate profile for $T<T^{\ast}\ $and the n-mer
profile for $T>T^{\ast}$. \ The fraction of particles in dimers, trimers,
and tetramers is also plotted. \ }
\label{melt2}
\end{figure}

Particles in the aggregate are pinned down by their nearest neighbor bonds,
so we do not consider their translational entropy. $\ $As a result the
chemical potential is simply $\mu_{\infty}=\epsilon_{\infty}$. \
Equilibrating the chemical potential of the monomer in solution and in the
aggregate we have: 
\begin{align}
T\log\left( c\right) & =\epsilon_{\infty} \\
\epsilon_{\infty} & =\epsilon_{AB}-T\log\left( \gamma_{\infty}\right)
\end{align}
Here $\gamma_{\infty}\simeq1$ is the configurational entropy of the branched
aggregate, per particle.

The concentration of particles in the aggregate $c_{\infty}$ is the the
total concentration minus the n-mer concentration. \ Here $c_{1}=c_{A}+c_{B}$
is the total monomer concentration, $c_{2}=c_{AB}$ is the total dimer
concentration, etc. \ 
\begin{equation}
c_{\infty}\approx c_{tot}-c_{1}-c_{2}-c_{3}-c_{4}
\end{equation}
Expressed in terms of $\widetilde{\epsilon}_{AB}$ and the fraction of solid
angle available to particles in the aggregate $\gamma_{\infty}=\frac {%
\Omega_{\infty}}{4\pi}$ we have: 
\begin{equation}
f_{\infty}=\frac{1}{\gamma_{\infty}}\exp\left[ \frac{\widetilde{\epsilon }%
_{AB}}{T}\right]  \label{cluster}
\end{equation}

The transition from dimers, trimers, etc. to the aggregation behavior is the
temperature $T^{\ast}$ at which $f_{\infty}(T^{\ast})$ is a solution to eq. %
\ref{polynomial}. \ In words, $T^{\ast}$ is the temperature at which the
aggregate has a non-zero volume fraction. \ The fraction of unbound
particles for these colloidal assemblies will be governed by eq. \ref%
{cluster} for $T<T^{\ast}$ and eq. \ref{polynomial} for $T>T^{\ast}$. \ As
claimed, we can simply relate the unbound fraction to $\widetilde{\epsilon}%
_{AB}$ for both n-mers and the aggregate. \ 

\subsection{Comparison to the experiment}

Let's consider the experimental scheme of Chaikin et al\cite{chaikin}. \ In
the experiment, $R=.5\mu m$ polystyrene beads were grafted with ds DNA
linkers of length $L\simeq20nm$. \ The 11 end bases of the $A$ and $B$ type
particles were single stranded and complementary. \ We have already
determined the bridging probability in this scenario(see section II.B). \ In
the experiment\cite{chaikin} a polymer brush is also grafted onto the
particle surface, which will have the effect of preferentially orienting the
rods normal to the surface(See Figure \ref{hybridization}). \ This
confinement of the linker DNA can be incorporated quite easily into our
results for $\Delta\widetilde{G}$ and $\left\langle N\right\rangle $. \
Returning to section II.B, when integrating over linker conformations we
simply confine each rigid rod to a cone of opening angle $2\alpha$. \ The
upper bound for the polar integration is now $\alpha$ as opposed to $\pi$. \ 
\begin{equation}
\Delta\widetilde{G}_{A}\simeq\Delta G_{A}+T\log\left[ 4\pi L^{3}c_{o}
(1-\cos\alpha)^{2}\right]
\end{equation}

The alignment effect should also be taken into account when calculating $%
\left\langle N\right\rangle $. \ If the particles are separated by less than 
$2L\cos\alpha$ the end sequences will be unable to hybridize. \ Following
the same steps as before, the lower bound for the $h^{^{\prime}}$
integration is now $L\cos\alpha$ as opposed to $0$. \ 
\begin{align}
\left\langle N\right\rangle & =8\pi^{2}\sigma^{2}R \int_{L\cos\alpha}^{L}
(L^{2}-h^{^{\prime}2})dh^{^{\prime}} \\
& =\frac{16}{3}\pi^{2}\sigma^{2}RL^{3}\left[ 1+\frac{\cos\alpha}{2}(\cos
^{2}\alpha-3)\right]
\end{align}
In the absence of the brush, and at sufficiently low linker grafting density 
$\sigma$, the alignment effect could be removed by setting $\alpha=\frac{\pi 
}{2}$, in which case we recover our previous results. Since the polymer
brush is stiff, it also imposes a minimum separation of $2h$ between
particles, where $h$ is the height of the brush. \ As a result, in the
expression for $\epsilon_{AB}$ we can approximate the radial flexibility of
the $AB$ bond as $\delta\simeq L-h$. \ 
\begin{figure}[h]
\begin{center}
\includegraphics[
height=3.2206in,
width=4.28in
]
{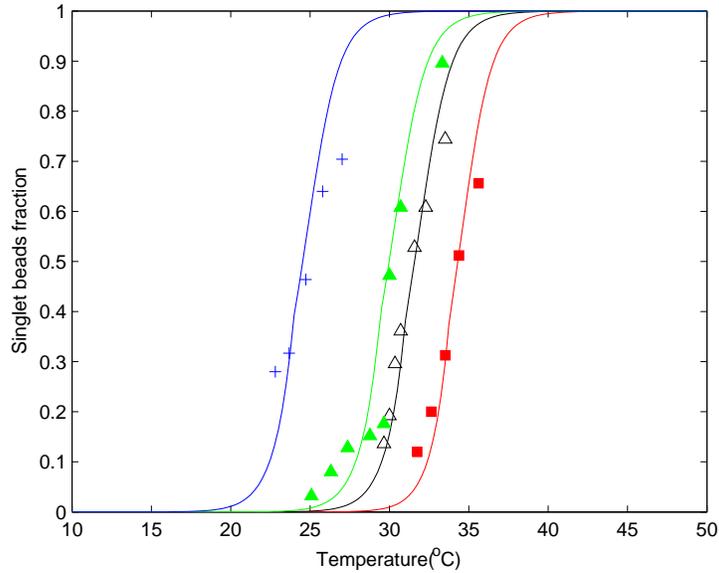}
\end{center}
\caption{(Color Online) Comparison of the melting curves $f(T)$ determined
by our model to the experimental data of Chaikin et al(See Fig.2 in 
\protect\cite{chaikin}). \ The four data sets are for the four different
polymer brushes used. \ For the model fits we find that $\left\langle
N\right\rangle _{2+}$=2.01 for crosses, 2.07 for solid triangles, 2.13 for
empty triangles, and 2.35 for squares. \ }
\label{chaikinplot}
\end{figure}

We have now related the free energy $\epsilon_{AB}$ to the known
thermodynamic parameters of DNA($\Delta G=\Delta H-T\Delta S$, $\Delta
H=-77.2\frac {kcal}{mol}$ and $\Delta S=-227.8\frac{cal}{molK}$), and the
properties of linker DNA chains attached to the particles(grafting density $%
\sigma \simeq3\times10^{3}\frac{DNA}{\mu m^{2}}$ and linker length $%
L\simeq20nm$). \ The height of the polymer brush is $h=13\pm5nm$\cite%
{chaikin}. \ In fitting the experimental data we have taken the average
value $\left\langle h\right\rangle =13nm$. \ Changing $h$ within these
bounds does not have a major effect on the melting curves. \ As a result
there is one free parameter in the model, the confinement angle $\alpha$. \
This angle determines $\left\langle N\right\rangle $ and $\Delta\widetilde{G}
$, which in turn determine $\widetilde{\epsilon}_{AB}$, and finally the
melting profile $f$. \ 

With some minor modifications we can also analyze the "tail to tail"
hybridization mode in a recent experiment of Mirkin et al\cite{meltingprop}.
\ In this experiment, $R=6.5nm$ gold nanoparticles were chemically
functionalized with ss DNA linkers. \ The last $15$ bases on the markers for
particles of type $A$ and $B$ were chosen to be complementary to a $30$ base
ss DNA linker. \ Since the strands are not ligated after hybridization, the
experimental pictures are similar. \ 
\begin{figure}[h]
\begin{center}
\includegraphics[
height=3.218in,
width=4.2774in
]
{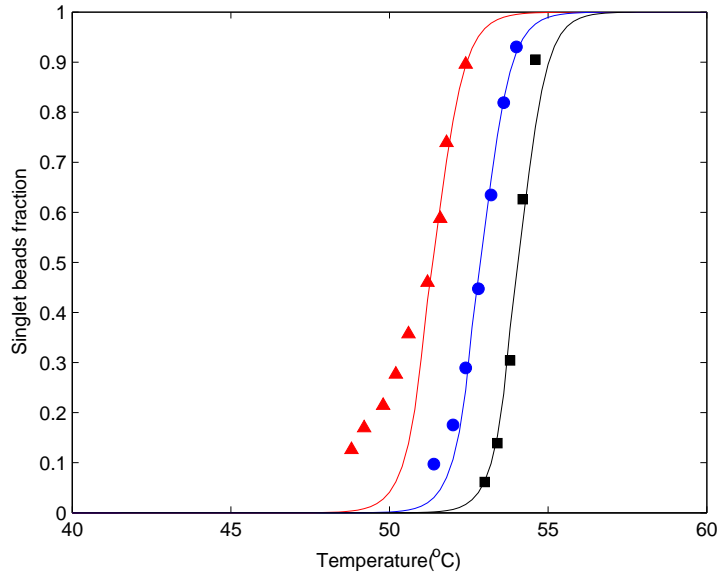}
\end{center}
\caption{(Color Online) The effect of the linker DNA grafting density $%
\protect\sigma$ on the melting profile$\ f(T)$. \ The results of the model
are compared with experimental data in \protect\cite{meltingprop}. The three
data sets represent grafting densities of 100\%(squares), 50\%(circles), and
33\%(triangles) for which $\left\langle N\right\rangle _{2+}$=2.32, 2.16,
and 2.05 respectively. \ }
\label{mirkinplot}
\end{figure}

The unhybridized portion of the ss DNA linker simply serves as a spacer, and
the hybridized portions become ds DNA, which we can again treat as rigid
rods(see section II.B). \ This experiment is done without the addition of a
polymer brush, but the grafting density is two orders of magnitude larger
than the experiment of Chaikin et al. \ As a result, there is still an
entropic repulsion\cite{morphology} associated with compressing the
particles below separation $2h$. \ Here $h$ could loosely be interpreted as
the radius of gyration of the unhybridized portion of the linker. \ Despite
the fact that $L\sim R$, our planar calculation of $\Delta\widetilde{G}$
provides a good fit to the experimental data. \ The other major difference
is that now the attraction between particles is mediated by an additional
DNA linker. 
\begin{equation}
\Delta G=\Delta G_{A}+\Delta G_{B}-T\log\left[ \frac{c_{link}}{c_{o}}\right]
\end{equation}

The first term is the contribution to the free energy from the hybridization
of the end sequence on linker $A$ to the complementary portion of the 30
base ss linker. \ The hybridization free energies $\Delta G_{A}$ and $\Delta
G_{B}$ were calculated with the DINAMelt web server\cite{dinamelt}. \ The
last term is the contribution to the free energy from the translational
entropy of the additional linker DNA, with $c_{link}$ the additional linker
concentration. \ This highlights some incorrect assumptions of the
thermodynamic melting model\cite{meltingprop}, where the two hybridization
free energies were not calculated separately, and the translational entropy
of the additional linker DNA\ was ignored. \ By introducing dilutent strands
to the system, one can probe the effect of the linker grafting density $%
\sigma$ on the melting properties of the assembly(See Figure 2B in \cite%
{meltingprop}). \ The agreement between the experimental data and our theory
is good, except at small $f$ values. \ This is not surprising, since
comparing the two requires relating the measurement of optical extinction to
the unbound fraction $f$. \ This is a nontrivial matter when dealing with
aggregation, which corresponds to the small $f$ regime. \ 

\section{Conclusion}

We have developed a statistical mechanical description of aggregation and
melting in DNA-mediated colloidal systems. \ First we obtained a general
result for two-particle binding energy in terms of DNA hybridization free
energy $\Delta G$, and two model--dependent parameters: the average number
of available bridges $\left\langle N\right\rangle $ and the effective DNA
concentration $c_{eff}$. \ We have also shown how these parameters can be
calculated for a particular bridging scheme. In our discussion we have
explicitly taken into account the partial ergodicity of the problem related
to slow binding-unbinding dynamics.

In the second part of the paper it was demonstrated that the fractions of
dimers, trimers and other clusters, including the infinite aggregate, are
universal functions of a parameter $\ \widetilde{\epsilon }_{AB}/T=\epsilon
_{AB}/T-\log \left[ c_{tot}/2c_{0}\right] $. \ We have applied the results
of our theory to a particular scheme when the DNA bridge is made of two
freely jointed rods (ds DNA). The obtained melting curves are in excellent
agreement with two types of experiments, done with particles of nanometer
and micron sizes. Furthermore, our analysis of the experimental data give an
additional insight into microscopic physics of DNA bridging in these
systems: it was shown that the experiments cannot be explained without
introduction of angular localization of linker ds DNA. \ The corresponding
localization angle $\alpha $ is the only fitting parameter of the model,
which allows one to fit both the position and width of the observed melting
curves.

There are several manifestations of the greater predictive power of our
statistical mechanics approach, compared to the earlier more
phenomenological models. \ First, once $\alpha $ is determined for a
particular system, our theory allows one to calculate the melting behavior
for an alternative choice of DNA linker sequences. \ Second, if the
resulting clusters are separated, for example in a density gradient tube,
the relative abundance of dimers, trimers, and tetramers can be compared to
the values determined from the theory. \ 

Finally, the theory predicts aging of the colloidal structures, one
experimental signature for which is hysteresis of the melting curves. \ Such
an experiment proceeds by preparing a system above the melting temperature,
and measuring the unbound fraction of colloids as the temperature is
lowered. \ The system is allowed to remain in this cooled state for a very
long time, perhaps months, during which multiple DNA bridges break and
reform. \ During this time the colloids relax into a more favorable
orientation state, including states which are not accessible by simply
rotating about the contact point formed by the first DNA bridge between
particles. \ This favorable orientation state is characterized by an average
number of DNA\ bridges $\left\langle N\right\rangle $ greater than what we
calculate in the partially ergodic regime. \ If the unbound fraction is then
measured as the temperature is increased, the melting curve will shift to a
higher temperature, consistent with a larger value of $\left\langle
N\right\rangle $. \ 

\section{Acknowledgements}

This work was supported by the ACS Petroleum Research Fund (grant PRF
\#44181-AC10), and by the Michigan Center for Theoretical Physics (award \#
MCTP 06-03). \ 

\bibliographystyle{achemso}
\bibliography{acompat,dna}

\end{document}